\documentclass{PoS}

\usepackage{lineno}

\title{Following up GW Alerts With MAGIC: The Third LIGO/Virgo Observation Run}

\ShortTitle{Following up GW Alerts With MAGIC: The Third LIGO/Virgo Observation Run}

\author{\speaker{Davide Miceli}$^1$, Lucio Angelo Antonelli$^2$, Josefa Becerra Gonzalez$^3$, Alessio Berti$^4$, \u{Z}eljka Bo\u{s}njak$^5$, Stefano Covino$^6$, Barbara De Lotto$^1$, Francesca Del Puppo$^7$, Susumu Inoue$^8$, Francesco Longo$^7$, Elena Moretti$^9$, Lara Nava$^6$, Barbara Patricelli$^{10,2}$, Antonio Stamerra$^2$ for the MAGIC Collaboration\footnote{\texttt{https://magic.mpp.mpg.de/} For collaboration list see PoS(ICRC2019)1177.} \\
	\llap{$^1$}University of Udine and INFN Trieste, Udine, Italy\\
	\llap{$^2$}INAF, Rome, Italy \\
	\llap{$^3$}IAC and Universidad de La Laguna, Tenerife, Spain \\
	\llap{$^4$}University of Torino and INFN Torino, Torino, Italy \\
	\llap{$^5$}Faculty of Electrical Engineering and Computing, University of Zagreb, Zagreb, Croatia \\
	\llap{$^6$}INAF, Osservatorio Astronomico di Brera, Merate, Italy \\
	\llap{$^7$}University and INFN Trieste, Trieste, Italy \\
	\llap{$^8$}RIKEN, Wako, Saitama, Japan \\
	\llap{$^9$}IFAE-BIST, Bellaterra (Barcelona), Spain \\
	\llap{$^{10}$}University and INFN Pisa, Pisa, Italy \\
	E-mail: \email{davide.miceli@ts.infn.it}}

\abstract{The first two LIGO and Virgo observation runs have been important milestones in the gravitational wave (GW) field,
	thanks to the detection of GW signals from ten binary black hole systems and a binary neutron star system \cite{abb1,abb2}.
	In order to fully characterize the emitting source, the remnant object and its environment, electromagnetic
	follow-up observations at different wavelengths are essential, as learned from the
	GW170817/GRB170817A case. Given the quite large localization uncertainties provided by interferometers, the main challenge faced by facilities with a narrow field of view (e.g. Imaging
	Atmospheric Cherenkov Telescopes, IACTs) is
	to setup a suitable follow-up strategy in order to observe sky regions with the highest probability to
	host the electromagnetic (EM) counterpart of the GW signal. As member of the EM follow-up community, the MAGIC collaboration
	joined this effort in 2014. As the third observation run (O3) is currently ongoing, where both LIGO and Virgo are
	expected to have much better sensitivities, MAGIC is refining its follow-up strategy to maximize the chances
	of observing the EM counterparts as soon as possible. In this contribution we will describe this strategy, focusing
	on the different observation cases, which mainly depends on the information available from both GW and EM
	partner facilities.}

\FullConference{36th International Cosmic Ray Conference -ICRC2019-\\
	July 24th - August 1st, 2019\\
	Madison, WI, U.S.A.}

\begin{document}

\section{Introduction}
The LIGO Scientific Collaboration and Virgo Collaboration (LVC) first two scientific runs, named O1 and O2, have provided exciting scientific discoveries and unprecedented results in the emerging field of gravitational wave (GW) astronomy. The O1 run led to the first direct detections of gravitational waves \cite{abb3,abb4}. These signals were connected with the coalescence of binary systems composed by two stellar mass black holes (BH). In the O2 run seven signals from BH/BH systems were detected and the first gravitational signal (GW170817) from the coalescence of a binary system formed by two neutron stars (NS) \cite{abb2} was discovered. The nearly simultaneously detection of a short gamma-ray burst (GRB) GRB170817A by Fermi and Integral satellites \cite{fermi1,int1} and the following discovery of its optical counterpart, called AT2017gfo \cite{at2017gfo}, have brought to an extraordinary follow-up campaign. This event is the first remarkable example of a multi-messenger campaign in which EM, GW and neutrino facilities instrument are involved, leading to the first ever detection of EM counterpart of a GW event. The O3 LVC scientific run is actually ongoing. Considering the upgraded sensitivity, an increasing number of events from binary system containing NSs or stellar-mass BHs with better localization is expected (up to $\sim$ few BH/BH per week, up to one NS/NS per month) \cite{ligo2}. The transient EM emission from binary systems containing at least one neutron star (NS/NS or NS/BH) is extremely interesting for the very-high energy (VHE) (E> 100 GeV) community, at early time as well as at late epoch. Short GRBs are indeed indicated as the most probable EM counterparts of binary mergers. In addition, the unbounded ejecta from binary mergers could be able to produce at late times shocks which may also generate a detectable VHE emission \cite{murase}. The scenarios and the energetic impact of these environments are still under debate but it is definitely worth investigating them. The MAGIC telescopes, as Imaging Atmospheric Cherenkov Telescope (IACT), have an important challenge to face and overcome. Considering the broad source localization provided by interferometers (from few tens up to few hundreds square degrees) and the narrow field of view of IACT instruments, it is mandatory to arrange an appropriate follow-up strategy to observe the sky regions where the probability to find EM counterparts is higher. In this contribution we will show the procedure developed for following-up GW events with MAGIC considering different observation cases.

\section{The MAGIC Automatic Alert System}
\begin{figure}
	\begin{center}
	\includegraphics[width=\textwidth]{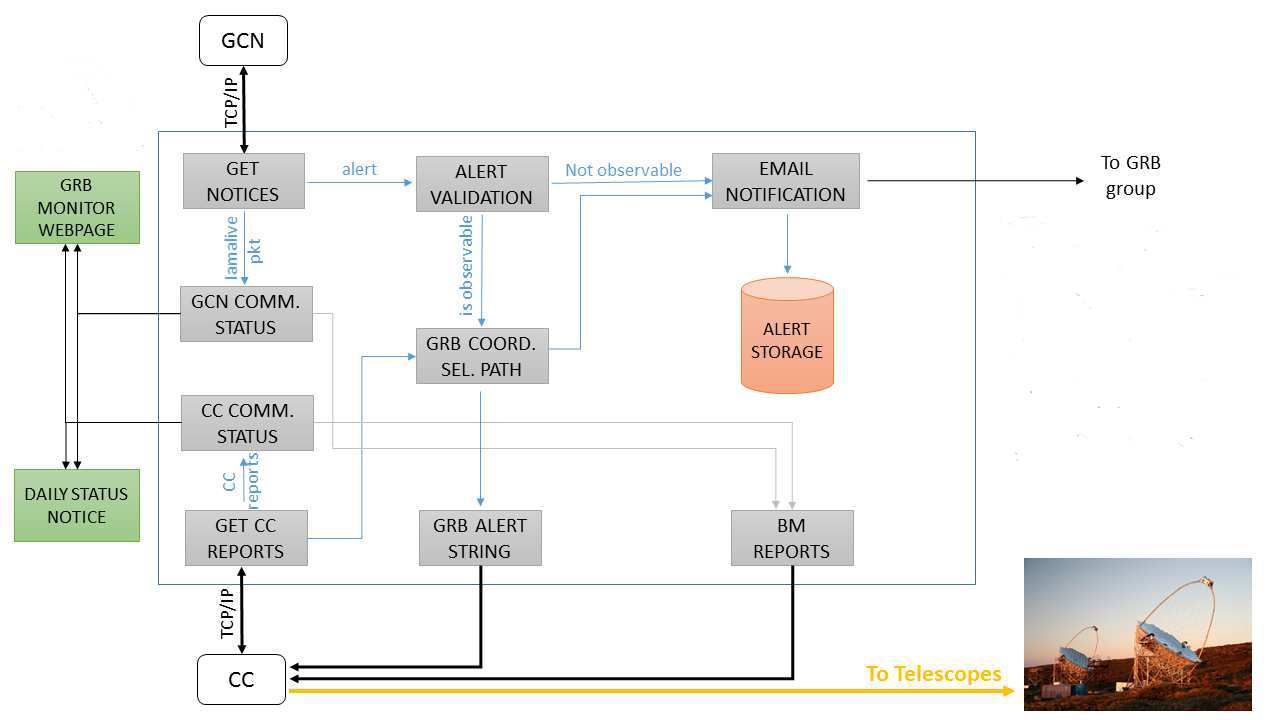}
	\caption{The main operations of the MAGIC Automatic Alert System. Credit: A. Carosi}
	\label{fig1}
	\end{center}
\end{figure}

MAGIC is a stereoscopic system of two Cherenkov Telescopes of 17 m diameter. They are located at the Roque de los Muchachos Observatory on the Canary Island of La Palma, Spain (28.8$^\circ$ N, 17.8$^\circ$ W) at an altitude of $\sim$ 2200 m a.s.l. \cite{magic1}. This instrument work in the VHE band using the gamma-ray indirect detection technique called Imaging Atmospheric Cherenkov technique. MAGIC can observe gamma-ray sources with an energy threshold of $\sim$ 50 GeV at zenith (or $\sim$ 30 GeV using the so called \textit{sum trigger}) and with a sensitivity at the level of 0.7$\%$ of the Crab Nebula flux above 220 GeV in 50 hours \cite{magic2}. Thanks to specific hardware configurations and a dedicated data analysis, MAGIC can also observe in moonlight conditions with good sensitivity \cite{magic3}. Moreover, with its lightweighted carbon-fiber structure, MAGIC can perform fast follow-up of transient sources and repoint every position in the sky very rapidly (25 s for a 180$^{\circ}$ rotation). This fast repositioning mode is extremely useful for follow-up of transient sources like Gamma-Ray Bursts, neutrino alerts, Fast Radio Bursts (FRBs) and counterparts of GW events. Considering the small field of view of the instrument (3.5$^{\circ}$), this responsive system was built up in order to rapidly react to external alerts released by other facilities through the Gamma-Ray Bursts Coordinates Network (GCN)\footnote{https://gcn.gsfc.nasa.gov/}. MAGIC can receive and handle these alerts thanks to an Automatic Alert System (AAS). A schematic representation of the structure is shown in Figure \ref{fig1}. The MAGIC AAS is a multi-threaded C program responsible for the communication with the GCN and the Central Control software of the MAGIC telescopes. The alerts coming from the GCN actually consist of two parts: the notices and the circulars. The former ones are sent through TCP/IP socket in form of binary packets and, thanks to their standard format, they can be automatically processed by the MAGIC AAS. These notices contain some information on the triggered events. The latter instead are text-readable messages sent via e-mail to inform the community on preliminary status and results from follow-up of the transient sources. In case of an alert coming from a GW candidate event, the MAGIC AAS also receives via internal e-mails the EM follow-up reports from the EM partners. The relevant information, namely time, pipeline, significance, neighbors, sky maps, is shared within the EM follow-up community through the GraceDB hub. In addition, the 3-D sky map with a posteriori mean luminosity distance \cite{gw1}, the probability that the less massive companion has a source-frame mass < 3 M$_\odot$ and the probability that the system ejected a significant amount of NS material are given.

The main actions of the MAGIC AAS can be summarized as:
\begin{itemize}
	\item acquire and verify the alerts coming from the GCN
	\item interpret the information contained in the alerts, especially the coordinates of the target
	\item according to some predefined observational criteria, check the visibility of the target from the MAGIC site
	\item take care of the automatic repointing procedure, if the target is observable.
\end{itemize}
The fast slewing of the telescopes is managed by the automatic procedure, also responsible to arrange the relevant subsystems for a correct data taking. This system was previously built for GRBs and neutrino events. For these transient events a specific target position in the sky is given in the alert so that the telescopes know immediately exactly where is the source to point. Instead, for GWs events, the alerts have a URL to an all-sky map where every pixel contains the probability value that a GW signal comes from that specific position \cite{ligo1}. So, a dedicated GW follow-up procedure is needed to process in a proper way the information obtained from this kind of alerts.

\section{GW-follow up procedure within MAGIC}

During O1 run the MAGIC collaboration has performed the first ever follow-up of a GW candidate (GW151226 \cite{abb4}) by a Cherenkov Telescope \cite{barbara}. Four different positions were manually selected and observed covering each a region of 2.5$^\circ$ x 2.5$^\circ$ of the skymap provided by LIGO. The positions were chosen according to few criteria like visibility, observations by other EM instruments, overlap with existing catalogs and using the LIGO localization skymap.
During O2 LVC scientific run, reacting to the GW event G299232, MAGIC observed three target positions related to a neutrino-muon candidate by IceCube, an optical transient and an unidentified transient detected by Swift/UVOT \cite{magicgcn2}. Moreover, we observed an optical transient then recognized as a SN-Ib from
CNEOST triggered by the GW event G298396. In addition, MAGIC has also participated in the observational campaign of AT2017gfo. MAGIC observations of the source lasted for a total amount of $\sim$ 9.2 hours from 15th January to 14th June 2018.
These past observations showed us that to follow-up rapidly the most promising GW events, MAGIC need an upgraded and automatic procedure to take care of the information given by the other facilities and investigate all possible interesting scenarios in the VHE domain. In particular, the implementation of a tiling strategy, namely a script to automatically define the pointing sequence, position and duration is mandatory. Moreover, the O2 scientific run of LVC has proven that EM counterparts of GW events can be detected in a wide energy range \cite{fermi1,int1,at2017gfo}. Therefore, also the strategy in case of observation at late time must be upgraded and standardized. For these reasons, a semi-automatic observational strategy has been developed for the MAGIC follow-up. We consider some possible observational cases of interest:
\begin{itemize}
	\item \textbf{Autonomous MAGIC follow-up}: this follow-up strategy will be applied in case a valid alert, in terms of visibility, zenith angle and telescope operational status, is received by MAGIC AAS and the GW candidate is quite well localized ($\sim$ 10$^\circ$ at 90$\%$ CL). The starting point is given by the LIGO localization skymap. From this map, the visibility of the sky positions from the MAGIC site at different times is computed. Then, the selected sky localizations with good visibility will be scanned with an ordered list of pointings. This scan list will take into account the galaxies included in the available catalogues \cite{gw2} or following the strategy proposed in \cite{gw3}. Finally, the telescopes will be repositioned to the target positions using the automatic procedure, described in the previous section, through the AAS.
	\item \textbf{Follow-up of identified transients}: in this case a transient is detected and localized by other facilities. The position will be pointed following the alert of the EM instrument released through standard channels (ATels, GCN). In case this happens during the night the standard automatic follow-up procedure will be followed, otherwise a dedicated observation will be scheduled.
\end{itemize}
In addition, also observations of possible delayed emission from GW events can be planned accordingly with the information that comes from other facilities. A set of monitoring observations can also be scheduled to optimize the time windows. This strategy was applied for the observations of the optical counterpart of the GW event GW170817, AT2017gfo, performed by MAGIC in the early 2018.

\section{Conclusion and Outlook}
The field of GW astronomy has gained in the last years several exciting new discoveries and detections that have pushed it into a new era. Moreover, the connection between an EM counterpart and a GW event from a BNS merger has generated an unprecedented follow-up multi-messenger campaign. Also MAGIC has participated in this campaign with the observation of AT2017gfo. The ongoing LIGO/Virgo observational run (O3), thanks to the improved sensitivity of the instruments, will provide many more GW events than before. Moreover, during the last part of O3 run (before October 2019), another GW interferometer, named KAGRA, is expected to start observations \cite{kagra}. It will be helpful to further increase the accurancy on the localization of GW source as well as the calculation of its main parameters. The MAGIC collaboration has refined a GW events follow-up strategy in order to investigate the most interesting ones, taking into account different observational cases. This strategy, hopefully, may lead to the detection of EM counterparts, whether GRB-like counterparts, in the first minutes following the candidate events, or, weeks/months later, VHE delayed emission.

\acknowledgments We would like to thank the Instituto de Astrof\'{\i}sica de Canarias for the excellent working conditions at the Observatorio del Roque de los Muchachos in La Palma. The financial support of the German BMBF and MPG, the Italian INFN and INAF, the Swiss National Fund SNF, the ERDF under the Spanish MINECO (FPA2017-87859-P, FPA2017-85668-P, FPA2017-82729-C6-2-R, FPA2017-82729-C6-6-R, FPA2017-82729-C6-5-R, AYA2015-71042-P, AYA2016-76012-C3-1-P, ESP2017-87055-C2-2-P, FPA2017‐90566‐REDC), the Indian Department of Atomic Energy, the Japanese JSPS and MEXT, the Bulgarian Ministry of Education and Science, National RI Roadmap Project DO1-153/28.08.2018 and the Academy of Finland grant nr. 320045 is gratefully acknowledged. This work was also supported by the Spanish Centro de Excelencia "Severo Ochoa" SEV-2016-0588 and SEV-2015-0548, and Unidad de Excelencia "Mar\'{\i}a de Maeztu" MDM-2014-0369, by the Croatian Science Foundation (HrZZ) Project IP-2016-06-9782 and the University of Rijeka Project 13.12.1.3.02, by the DFG Collaborative Research Centers SFB823/C4 and SFB876/C3, the Polish National Research Centre grant UMO-2016/22/M/ST9/00382 and by the Brazilian MCTIC, CNPq and FAPERJ.

\end{document}